\documentclass[a4paper,11pt]{article}
\usepackage{jinstpub} 
\usepackage{lineno}

\proceeding{ACAT 2025: 23rd International Workshop on Advanced Computing and Analysis Techniques in Physics Research\\
September 8th-12th, 2025\\
Hamburg, Germany}

\title{\boldmath Hardware-Aware Design of a GNN-Based Hit Filtering Algorithm for the Belle II Level-1 Trigger}

\author[a,1]{G. Heine,\note{Corresponding author.}}
\author[b]{F. Mayer,}
\author[b]{M. Neu,}
\author[b]{J. Becker,}
\author[a]{T. Ferber}
\affiliation[a]{Institute of Experimental Particle Physics, Karlsruhe Institute of Technology,\\Wolfgang-Gaede-Straße 1, 76131 Karlsruhe, Germany}
\affiliation[b]{Institute for Information Processing Technology, Karlsruhe Institute of Technology,\\Engesserstraße 5, 76131 Karlsruhe, Germany}

\emailAdd{greta.heine@kit.edu}

\abstract{The Belle~II experiment operates at high luminosity, where an increasing beam-induced background imposes stringent demands on the hardware Level-1 trigger system, which must operate under tight latency and bandwidth constraints.
To achieve online data reduction within the Level-1 trigger system, we have developed a hit-filtering algorithm based on the lightweight Interaction Network architecture.
In this work, we present a hardware-aware model-compression workflow for this  hit-filtering algorithm targeting deployment on FPGA devices within the Belle~II trigger system.
The network is adapted to the detector and trigger conditions through model-size and graph-size reduction, low-precision (4\,bit) fixed-point arithmetic, and unstructured pruning.
We assess the resulting design using the total number of bit operations as a hardware-aware computational complexity metric.
Using this metric, we identify a configuration that decreases this cost by more than two orders of magnitude relative to the full-precision reference implementation.
This reduction is achieved while preserving performance close to the reference model in terms of hit efficiency and background rejection, as indicated by only a modest decrease in the AUC score from \num{97.4} to \num{96.8}, evaluated on Belle~II collision data.}


\newacronym{daq}{DAQ}{Data Acquisition}
\newacronym{l1}{L1}{Level-1}
\newacronym{hlt}{HLT}{High-Level Trigger}
\newacronym{fee}{FEE}{Front-End Electronics}
\newacronym{grl}{GRL}{Global Reconstruction Logic}
\newacronym{gdl}{GDL}{Global Decision Logic}
\newacronym{fpga}{FPGA}{Field-Programmable Gate Array}
\newacronym{cdc}{CDC}{Central Drift Chamber}
\newacronym{klm}{KLM}{K-Long and Muon detector}
\newacronym{ecl}{ECL}{Electromagnetic Calorimeter}
\newacronym{adc}{ADC}{Analog-to-Digital Converter}
\newacronym{tdc}{TDC}{Time-to-Digital Converter}
\newacronym{ut4}{UT4}{Universal Trigger Board 4}

\newacronym{in}{IN}{Interaction Network}
\newacronym{gnn}{GNN}{Graph Neural Network}
\newacronym{mva}{MVA}{Multi-Variate Analysis}
\newacronym{mlp}{MLP}{Multi-Layer Perceptron}
\newacronym{mc}{MC}{Monte Carlo}
\newacronym{roc}{ROC}{Receiver Operating Characteristic}
\newacronym{auc}{AUC}{Area Under the Curve}
\newacronym{bop}{BOP}{number of Bit Operation}
\newacronym{mac}{MAC}{multiply-accumulate}
\newacronym{basf2}{basf2}{Belle~II Analysis Software Framework}

\newacronym{lut}{LUT}{Look-Up Table}
\newacronym{ff}{FF}{Flip-Flop}
\newacronym{dsp}{DSP}{Digital Signal Processing}
\newacronym{pe}{PE}{Processing Element}
\newacronym{rtl}{RTL}{Register-Transfer Level}
\newacronym{fifo}{FIFO}{First-In-First-Out buffers}
\glsdisablehyper
\notoc

\begin{document}
\maketitle
\flushbottom
\glsresetall
\section{Introduction}
\label{sec:intro}
The Belle~II experiment \cite{Belle-II:2010dht} at the SuperKEKB $e^+e^-$ collider is designed to search for physics beyond the Standard Model via precision measurements of rare decays.
To handle substantial beam-induced background at high instantaneous luminosity, the \gls{daq} system employs a hardware \gls{l1} trigger that selects physics-relevant events in real-time, significantly reducing the data throughput for detector readout.
The \gls{l1} trigger must maintain a high efficiency for physics signals while operating within a strict \SI{5}{\micro\second} latency constraint and staying within the \gls{daq} limits \cite{Lai:2025gac, Yamada:2015xjy}. 
One important sub-trigger of the \gls{l1} trigger system is the \gls{cdc} trigger \cite{Taniguchi:2017not}, providing track information for the final trigger decision logic.
The increasing background directly impacts track reconstruction efficiency, purity, and precision~\cite{Aihara:2024zds,Lai:2025gac,Liu:2025iup}, necessitating fast and effective hit filtering prior to track finding. \\
\Glspl{gnn} have been studied for particle tracking and related reconstruction tasks in high-energy physics, as they can represent irregular detector geometry and local hit correlations in a graph structure \cite{DeZoort:2021rbj, ExaTrkX:2020nyf}.
Deploying such \glspl{gnn} in first-level trigger systems introduces additional constraints: the end-to-end inference latency, including graph construction, must be kept in the sub-microsecond range; the model must fit into limited \gls{fpga} resources; and physics performance in terms of track or cluster reconstruction accuracy needs to be preserved despite model compression and quantization \cite{Elabd:2021lgo, Neu:2023sfh}.
For the Belle~II \gls{l1} trigger system, the target hit-filtering application, planned to be distributed across 20 \gls{fpga} boards, is required to process up to \num{978} sense wires per \gls{cdc} sector within a sub-microsecond latency budget, providing hit-filter outputs for every \SI{32}{\mega\hertz} \gls{cdc} trigger clock cycle. \\
This work describes a software-hardware co-design procedure for a \gls{gnn}-based hit-filtering algorithm under these constraints. In contrast to previous work that primarily documented the final hardware implementation \cite{Heine:2025xit}, the present work focuses on the iterative design steps in which hardware limitations, fixed-point precision, and resource usage guide the development of the model architecture. \\
The remainder of this paper is organized as follows. \Cref{sec:gnn} describes the compression pipeline, starting from the full-precision \gls{gnn} hit-filtering model and its graph inputs, and then outlining the model-size reduction, quantization-aware training, and pruning steps used to obtain a deployable design. \Cref{sec:evaluation} reports the performance evaluation, considering both hit-filtering efficiency and background rejection, and the corresponding number of bit operations as a hardware-aware cost metric summarized in \cref{sec:conclusion}.
\section{Compression pipeline: from full-precision to online}
\label{sec:gnn}
\begin{figure}
    \centering
    \includegraphics[width=1\linewidth]{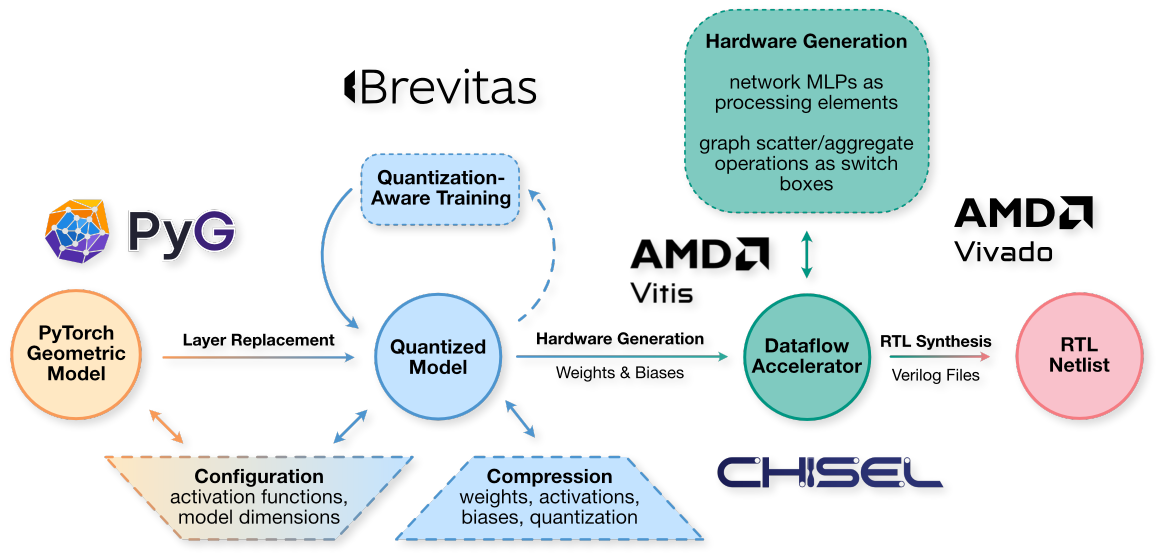}
    \caption{Our workflow for deploying a PyTorch~Geometric model on \acrshort{fpga}: starting from model configuration and layer replacement with Brevitas quantization layers, followed by model compression including quantization-aware training, hardware generation for the dataflow \gls{gnn} accelerator, and subsequent \acrshort{rtl} netlist creation for \gls{fpga} implementation.}
    \label{fig:workflow}
\end{figure}

In this section, we describe our workflow for deploying our \gls{gnn} hit-filtering model on \gls{fpga}, as illustrated in \cref{fig:workflow}.
Starting from a full-precision PyTorch~Geometric \cite{Fey:2019wpv} model described in~\autoref{sec:GNN_fullmodel}, size reduction, quantization-aware training and pruning are applied to obtain a compressed, low-precision network described in ~\autoref{sec:GNN_quantmodel}. 
From this compressed model, a hardware description for the dataflow \gls{gnn} accelerator is generated and subsequently synthesized into an \gls{rtl} netlist, which serves as the basis for the final \gls{fpga} implementation detailed in \cite{Heine:2025xit}.

\subsection{Full-precision baseline model}
\label{sec:GNN_fullmodel}
Our baseline model employs a modified Interaction Network \cite{Battaglia:2016jem} as a lightweight \gls{gnn} architecture, particularly suited to relationship modelling and pattern recognition.
The network consists of three sequential \gls{mlp} blocks: an edge block $R_1$ (edge feature update), a node block $O$ (node feature update), and a final edge block $R_2$ (final edge classification). 
For each edge the first edge block $R_1$ processes the edge features and associated incoming and outgoing node features $(x_{in}, x_{out}, e)$, producing updated edge features $\tilde{e}$. 
Updated edge features $\tilde{e}$ connected to each node are then aggregated via a max-scatter operation to $\bar{e}$, retaining the largest edge response per node. The subsequent node block $O$ takes $\bar{e}$ together with the node features $x$ and outputs updated node features $\tilde{x}$, which are passed, together with $\tilde{e}$, to the final edge block $R_2$ as $(\tilde{x}_{in}, \tilde{x}_{out}, \tilde{e})$. 
The one-dimensional edge score outputs $\tilde{\tilde{e}}$ of $R_2$ are aggregated per node by a mean-scatter operation, yielding node-level scores. 
A final sigmoid activation converts these node scores to node-wise probability outputs $\tilde{\tilde{x}}$ between 0 and 1. 
In the baseline model configuration, each \gls{mlp} block comprises three ReLu-activated hidden layers with dimensions [8,8,3], [8,8,3], and [8,8,1] respectively, resulting in 495 trainable parameters in total.

\paragraph{Graph inputs:} A graph representation is constructed from the detector hits by connecting each sense wire to pattern-based neighbour wires via bidirectional edges \cite{Neu:2023sfh}.
Node features $x_{in/out}$ encode the wire positions ($x$ and $y$ at the wire centre~\cite{Belle-II:2010dht}) as well as the \gls{adc} count sum per wire, while edge features $e$ capture spatial ($\Delta r$ radial and $\Delta \phi$ azimuthal) and temporal ($\Delta$TDC) differences between connected wires.

\paragraph{Training strategy:} For training, we designed a dataset covering a wide range of track signatures and background conditions, using both simulated events and Belle~II 2024 collision data. Details on the dataset, training strategy, and graph construction are provided in \cite{Heine:2025xit}. 

\subsection{Compressed quantized model}
\label{sec:GNN_quantmodel}
For deployment of the \gls{gnn} model on the target AMD Ultrascale XCVU190 device, network compression and careful adaptation to hardware constraints are required.
The final deployed configuration was found by applying the following compression and optimization steps while monitoring both model prediction accuracy and \gls{fpga} system resource utilization.
 
\paragraph{Model and graph size reduction:} First, the number of trainable parameters is reduced with respect to the full-precision model by decreasing the number of hidden layers per \gls{mlp} block from two to one and shrinking the remaining hidden layer size from eight to six neurons each, resulting in a reduction from 495 to 211 trainable parameters.
In addition, the graph size is reduced by switching from bidirectional to unidirectional edges, effectively halving the number of edges and thus the computational load.

\paragraph{4bit quantization:} To further compress the network and prepare it for fixed-point implementation, quantization-aware training is performed using Brevitas \cite{brevitas}. 
We replace the floating-point PyTorch network layers with the differentiable Brevitas quantized equivalents, using custom quantizers applying fixed-point quantization with floor rounding mode to the weights, biases, and activation functions of the model.
We add \gls{mlp} wrapping with quantized identity layers to enforce quantization boundaries for in- and outputs to avoid accumulation overflows. We use a mixed precision scheme to balance resource constraints with numerical stability due to accumulation effects: \SI{4}{\bit} inputs and \SI{4}{\bit} weights, \SI{6}{\bit} activations, \SI{16}{\bit} bias, \SI{8}{\bit} outputs.

\paragraph{Pruning:} In addition, we apply an iterative magnitude-based unstructured pruning, linearly increasing the weight sparsity every five training epochs up to a final sparsity of \SI{65}{\percent}. \\ \\
For hardware implementation, the exponential sigmoid output activation is removed post-training, since, during inference, classification decisions rely solely on thresholding the resulting score.

\section{Performance evaluation}
\begin{figure}[btp]
    \centering
    \begin{subfigure}{0.4170\textwidth}
        \centering
        \includegraphics[width=\textwidth]{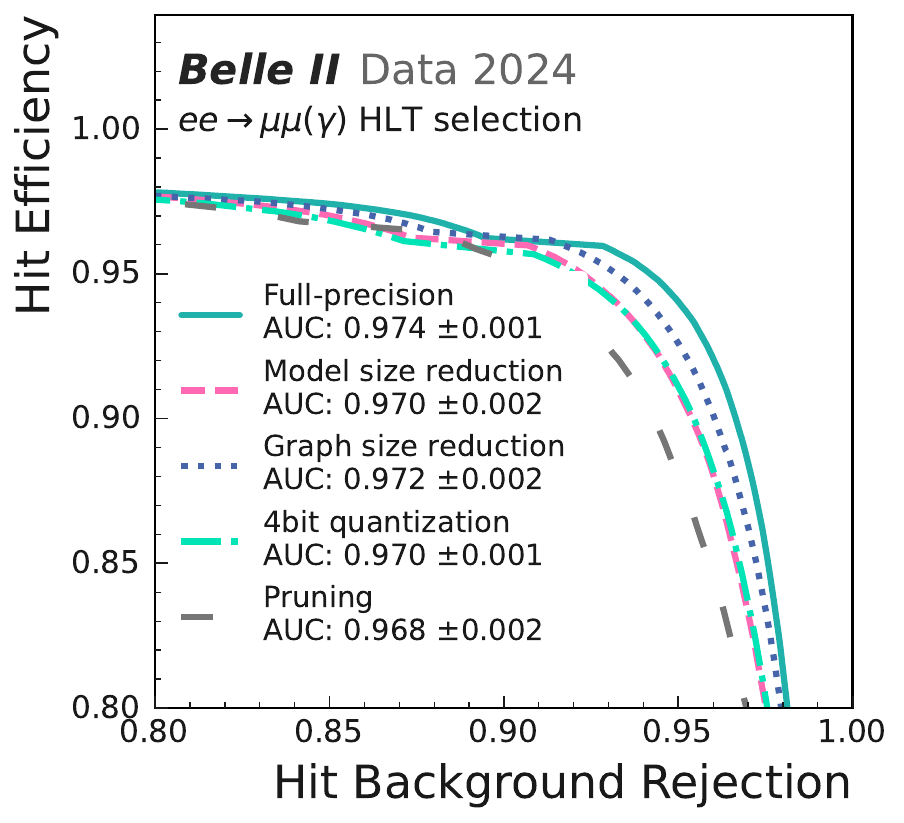} 
        \caption{ROC curves}
        \label{fig:roc}
    \end{subfigure}
    \hfill
    \begin{subfigure}{0.54\textwidth}
        \centering
        \includegraphics[width=\textwidth]{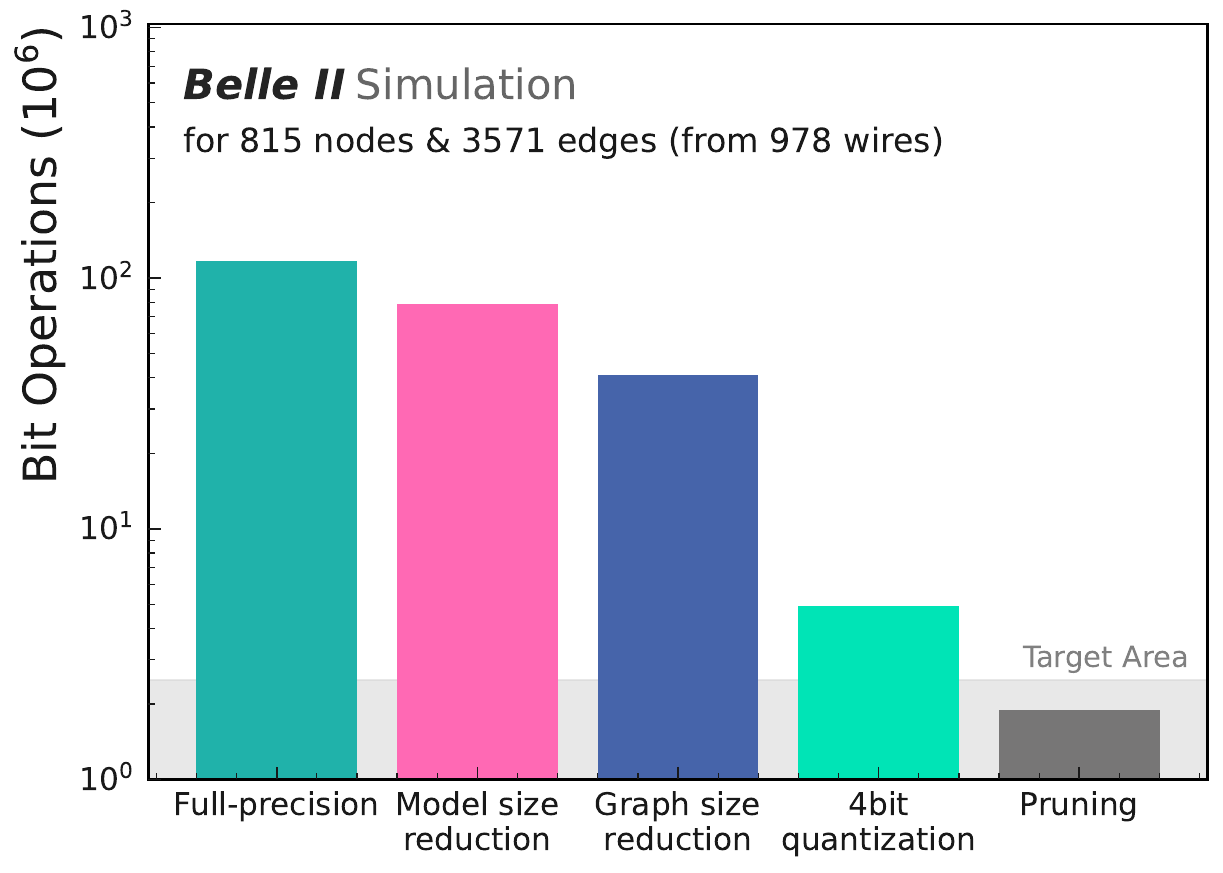}
        \caption{Number of bit operations}
        \label{fig:bops}
    \end{subfigure}
    \caption{\textbf{(\subref{fig:roc})} Hit-level classification \acrfull{roc} curves and \acrfull{auc} values for different network compression steps and \textbf{(\subref{fig:bops})} corresponding \glspl{bop} as a proxy for computational complexity towards \gls{fpga} implementation.}
    \label{fig:results}
\end{figure}

For the main compression steps between the offline full-precision reference and the final online model, as discussed in \cref{sec:gnn}, the \gls{roc} curves and the \glspl{bop} are shown in \cref{fig:results}.
The following configurations are considered, where each step includes all preceding modifications: \textbf{1) Full-precision} baseline model; \textbf{2) model size reduction} (of hidden depth and width) with 211 instead of 495 trainable parameters; \textbf{3) graph size reduction} by using unidirectional instead of bidirectional edges; \textbf{4) 4bit quantization} with 4bit weights and inputs including removal of the output sigmoid activation, which does not affect classification performance or \glspl{bop}; and \textbf{5) pruning} with \SI{65}{\percent} final unstructured weight sparsity.
\label{sec:evaluation}
 \noindent

\subsection{Hit filtering performance}
\label{subsec:roc}
The hit filtering performance shown in~\autoref{fig:roc} is evaluated in the form of \gls{roc} curves and associated \gls{auc} scores through an offline re-analysis of Belle~II data.
The test sample consists of hits from \num{1000} \gls{hlt}-selected $\mu\mu(\gamma)$ events\footnote{\gls{hlt} $\mu\mu(\gamma)$ events are selected by requiring two oppositely charged, back-to-back tracks, each carrying momentum greater than $\SI{0.5}{\giga\electronvolt/c}$ in the center-of-mass reference frame and matched to an \gls{ecl} cluster with energy below $\SI{0.5}{\giga\electronvolt}$. The total energy of clusters, including possible photons, must be below \SI{2}{\giga\electronvolt}~\cite{Kuhr:2018lps}.
} recorded at the end of 2024, corresponding to approximately \SI{2.5}{million} \gls{cdc} hits.
Due to stochastic variations between training runs (e.g., from random initialization), for each configuration, the reported values are the mean and standard deviation over five independent training runs, where in each run the best-performing model out of three runs is selected. 
Pre-selection cuts (\gls{adc}$\geq10$ and \gls{tdc} within \SI{500}{\nano\second} trigger time window) consistent with the default Belle~II \gls{l1} trigger setup are applied before network training and inference, limiting the maximum achievable hit efficiency (defined as the fraction of signal hits retained over all available signal hits) to \num{0.980}. 
Signal hits are defined as hits on sense wires connected to the \gls{l1} trigger that are associated with tracks found by the Belle~II offline reconstruction \cite{Kuhr:2018lps,the_belle_ii_collaboration_2025_16268234} passing a quality selection.\footnote{Track selection requires transverse momentum $p_\mathrm{T} > \SI{0.2}{\giga\electronvolt/c}$, total momentum $p > \SI{0.7}{\giga\electronvolt/c}$, longitudinal distance $|z_0| < \SI{15}{\centi\meter}$ and radial $|d_0| < \SI{15}{\centi\meter}$ distance from the interaction point, and at least 7 \gls{cdc} hits.} 
Background hits are defined as hits not associated with found tracks.
The background rejection rate is given by the fraction of such hits removed by the classifier. Hits matched to tracks failing quality criteria constitute $<\SI{5}{\percent}$ of signal and $<\SI{0.1}{\percent}$ of all hits and are excluded from performance metrics. \\ 
The full-precision model attains an \gls{auc} of \num{0.974\pm0.001} and a background rejection of \SI{94.2\pm0.4}{\percent} at a hit efficiency of \SI{95}{\percent}, whereas the final compressed and pruned model achieves an \gls{auc} of \num{0.968\pm0.002} with a background rejection of \SI{90.9\pm1.5}{\percent}. 
Thus, the cumulative degradation in hit-filtering performance introduced by the compression pipeline remains modest.

\subsection{Number of bit operations}
\label{subsec:ebops}
To characterize and compare the computational complexity of the different configurations, we estimate the number of \glspl{bop} per network layer, following~\cite{Baskin_2019}. 
For an \gls{mlp} layer with $N_w$ weights, $N_b$ biases, and $N_{in}$ inputs, and corresponding bit widths $b_w$, $b_b$, and $b_{in}$, the bit-operation count is 
\begin{equation}
    \text{BOPs}_{\text{layer}} = N_w b_w b_{in} + N_{b} b_{b} + N_w b_{acc},
\end{equation}
with accumulator width $b_{acc} = b_{in} + b_{w} + \log_2{N_{in}}$. 
The total number of \gls{mac} operations for the interaction network architecture on a graph of size $(N_\text{nodes}, N_\text{edges})$ is then
\begin{equation}
    \text{BOPs} = N_\text{edges} \cdot (\text{BOPs}_{R_1} + \text{BOPs}_{R_2}) + N_\text{nodes} \cdot (\text{BOPs}_{O} + \text{BOPs}_\text{aggr}),
\end{equation}
with $\text{BOPs}_\text{MLP} = \sum_{\text{layer} \in \text{MLP}} \text{BOPs}_\text{layer}$ for $ \text{MLP} \in \{R_1, O, R_2\}$.
In the following, we assume that the aggregation term is negligible compared to the dominant weight-multiplication terms, i.e., $\text{BOPs}_\text{aggr} \ll \text{BOPs}_\text{MLP}$, and we therefore omit it from the \gls{bop} calculation. \\
\Cref{fig:bops} shows the \glspl{bop} count for each compression step for athe largest \gls{cdc} sector with 978 sense wires.
Restricting the graphs to wires relevant to the \gls{l1} track trigger yields $N_\text{nodes}=\num{815}$ and $N_\text{edges}=\num{3571}$ for this sector. 
Based on preliminary hardware studies, we define a target range of \qtyrange[range-units = single]{1.0}{2.5}{MBOPs} per sector as compatible with the resource constraints of the AMD Ultrascale XCVU190.
While the \gls{bop} metric provides a useful, hardware-aware proxy for computational complexity, it remains a rough estimate and does not guarantee that a model configuration meets all resource and timing constraints on a specific \gls{fpga}.
To reflect this uncertainty, a target range is defined instead of a single threshold value. \\
The initial full-precision model with \SI{116.6}{MBOPs} exceeds this range by over two orders of magnitude, whereas the final compressed configuration achieves \SI{1.8}{MBOPs}, falling within the target range while maintaining near-baseline classification performance as discussed in \autoref{subsec:roc}. \\
The final configuration was validated using an out-of-context \gls{fpga} implementation targeting an AMD Ultrascale XCVU190 device, as detailed in \cite{Heine:2025xit}, which reports system resource utilization after synthesis, placement, and routing.
The implementation for a graph with 495 nodes and 2163 edges uses \SI{35.65}{\percent} of the available \glspl{lut}, \SI{29.75}{\percent} of the \glspl{ff}, no \glspl{dsp}, and achieves a total pipeline latency of  \SI{632.4}{\nano\second} at the system frequency of \SI{128}{\mega\hertz}, satisfying the sub-microsecond latency requirement.

\section{Conclusion}
\label{sec:conclusion}
This work presents a systematic workflow to compress a \gls{gnn}-based hit-filtering model for deployment in the Belle~II \gls{l1} trigger.
Starting from a full-precision PyTorch Geometric network, the model is compressed via architectural downsizing, mixed fixed-point quantization, and unstructured pruning.
To relate model changes to hardware utilization, we use the number of bit operations as a hardware-aware complexity metric.
The compression reduces the \glspl{bop} from \SIrange[range-units=single]{116.6}{1.8}{MBOPs} for the largest sector.
An out-of-context implementation on an AMD Ultrascale SCVU190 validates the final design, achieving sub-microsecond latency and moderate \gls{lut} and \gls{ff} utilization without \gls{dsp} usage.
Despite the substantial compression, the final model configuration preserves an \gls{auc} score that is close to that of the full-precision reference model (\num{0.974} vs. \num{0.968}) when evaluated on Belle~II 2024 collision data.


\appendix
\acknowledgments
We thank the Belle II trigger team for their support and feedback. We also thank T.~Koga for comments on the manuscript, and G.~D.~Pietro for help with the Belle~II software framework.
This work is funded by BMFTR ErUM-Pro 05H24VK1.


\bibliographystyle{JHEP}
\bibliography{biblio.bib}



\end{document}